\documentclass[fleqn,usenatbib]{mnras}

\usepackage{newtxtext,newtxmath}

\usepackage[T1]{fontenc}
\usepackage{ae,aecompl}

\usepackage{graphicx}	
\usepackage{amsmath}	
\usepackage{amssymb}	

\usepackage{xargs} 
\usepackage{amsmath}
\title[Polarimetry of {\it Swift} J2058+0516]{Polarimetry of relativistic tidal disruption event {\it Swift} J2058+0516}

\author[K. Wiersema et al.]{\parbox{\textwidth}{
K. Wiersema$^{1,2}$ \thanks{E-mail: K.Wiersema@warwick.ac.uk},  
A. B. Higgins$^{2}$, 
A. J. Levan$^{1,3}$, 
R. A. J. Eyles$^{2}$,
R. L. C. Starling$^{2}$, 
N. R. Tanvir$^{2}$,
S. B. Cenko$^{4,5}$, 
A. J. van der Horst$^{6,7}$,
B. P. Gompertz$^{1}$,
J. Greiner$^{8}$, 
D. R. Pasham$^{9}$
\vspace{0.4cm}}\\
$^{1}$ Department of Physics, University of Warwick, Coventry CV4 7AL, UK\\
$^{2}$ University of Leicester, University Road, Leicester LE1 7RH, UK\\
$^{3}$ Department of Astrophysics/IMAPP, Radboud University, Nijmegen, The Netherlands\\
$^{4}$ Astrophysics Science Division, NASA Goddard Space Flight Center, Mail Code 661, Greenbelt, MD 20771, USA\\
$^{5}$ Joint Space-Science Institute, University of Maryland, College Park, MD 20742, USA\\
$^{6}$ Department of Physics, The George Washington University, 725 21st Street NW, Washington, DC 20052, USA\\
$^{7}$ Astronomy, Physics, and Statistics Institute of Sciences (APSIS), The George Washington University, Washington, DC 20052, USA\\
$^{8}$ Max-Planck Institut f\"ur extraterrestrische Physik, 85740 Garching, 
 Giessenbachstr. 1, Germany\\
$^{9}$ MIT Kavli Institute for Astrophysics and Space Research, Cambridge, MA 02139\\
}

\date{Accepted XXX. Received YYY; in original form ZZZ}

\pubyear{2018}

\begin{document}
\label{firstpage}
\pagerange{\pageref{firstpage}--\pageref{lastpage}}
\maketitle

\begin{abstract}
A small fraction of candidate tidal disruption events (TDEs) show evidence of powerful relativistic jets, which are particularly pronounced at radio wavelengths, and likely  contribute non-thermal emission at a wide range of wavelengths. A non-thermal emission component can be diagnosed using linear polarimetry, even when the total received light is dominated by emission from an accretion disk or disk outflow. In this paper we present Very Large Telescope (VLT) measurements of the linear polarisation of the optical light of jetted TDE {\it Swift} J2058+0516. This is the second jetted TDE studied in this manner, after {\it Swift} J1644+57. We find evidence of non-zero optical linear polarisation, $P_{\rm V}\sim 8\%$, a level very similar to the near-infrared polarimetry of {\it Swift} J1644+57. These detections  provide an independent test of the emission mechanisms of the multiwavelength emission of jetted tidal disruption events.
\end{abstract}




\section{Introduction}
In recent years, a multitude of tidal disruption event candidates have been found in UV and optical widefield surveys, some of which also showing bright X-ray emission (e.g. \citealt{Komossa}). Generally, the optical and UV emission seems to follow a thermal spectrum, with temperatures in the region of $T\sim10^4$ K. 
A rare subset of TDEs have been shown to produce powerful relativistic jets.
To date, there are three firmly established relativistic tidal disruption events, {\it Swift} J1644$+$57 (e.g. \citealt{Bloom}; \citealt{Burrows}; \citealt{Levan}); {\it Swift} J2058$+$0516 (e.g. \citealt{Cenko}; \citealt{Pasham}) and {\it Swift} J1112.2$-$8238 (\citealt{Brown,Brownlate}). 
In all three cases, the sources showed bright, rapidly variable and long lasting X-ray emission (brighter than commonly seen in thermal spectrum TDEs), and a bright radio counterpart.  All three sources
were first identified in data from the Burst Alert Telescope (BAT),
a $\gamma$-ray instrument on board the Neil Gehrels {\em Swift} Observatory (hereafter {\em Swift}). Two of these sources have particularly good observational coverage: {\it Swift} J1644$+$57 and {\it Swift} J2058$+$0516. 
In both these cases, broadband spectral energy distributions (SEDs) were obtained, spanning from low frequency radio wavelengths all the way to high energy $\gamma$-rays. The two sources differ somewhat in their optical and infrared behaviour: {\it Swift} J1644$+$57 shows high levels of extinction (e.g. \citealt{Levan}; \citealt{Bloom}; \citealt{Burrows}; \citealt{Levan1644late}), making its optical and UV emission difficult to detect (\citealt{Levan1644late}). {\it Swift} J2058$+$0516 suffers much less from extinction, resides in a fainter host,and is at far higher redshift ($z=1.18$ vs $z=0.34$) enabling the optical and rest frame UV emission from the transient to be studied in greater detail. \cite{Pasham} show that the optical and infrared SED of {\it Swift} J2058$+$0516 can be described by a cooling blackbody-like spectrum  in the rest frame, with a relatively constant radius, qualitatively similar to a large number of TDEs that are not seen to have relativistic jets,
although significantly more luminous. 
The X-ray lightcurves of {\it Swift} J2058$+$0516 and {\it Swift} J1644$+$57 both show rapid variability at early times (e.g. \citealt{Pasham}; \citealt{Krolik}; \citealt{Saxton}), and a very steep and sudden brightness drop half a year after the TDE $\gamma$-ray trigger \citep{Pasham,Levan1644late}. The
fast variability may provide a way to measure black hole masses, if the steep drop-off
is associated with a transition from super-Eddington to sub-Eddington accretion regimes.

The detection rate of relativistic TDEs is low compared to the number of TDEs whose emission is generally thermal in the optical part of the spectrum (e.g. \citealt{SaxtonXray}), and for which there is no relativistic jet component readily visible in the SED (see e.g. \citealt{Auchettl}). To date, searches for non-thermal emission in thermal TDEs have largely focussed on deep radio and X-ray searches (e.g. \citealt{vanVelzen}), with a detection of a radio jet in at least one thermal TDE (\citealt{vanVelzenASASSN14li}; \citealt{Alexander}). Similar to black hole X-ray binaries, a weak jet component may also be found through linear polarimetry at optical and near-infrared wavelengths (see \citealt{Russell} and references therein), which makes polarimetry an important additional diagnostic tool, independent of SED and lightcurve models. Linear polarimetry has so far been limited to one jetted TDE ({\it Swift} J1644+57, \citealt{Wiersema1644}) and one TDE without jet detection (OGLE16aaa, \citealt{Higgins}). The infrared ($K_s$ band) imaging polarimetry of {\it Swift} J1644+57 showed a sizeable degree of linear polarisation ( $P\sim7.5 \pm 3.5\%$, measured after the initial steep decay phase and just prior a shallow decay phase, $\sim18$ days after trigger; see lightcurve in Fig.~3 of \cite{Wiersema1644}), which motivated the polarimetric observations of {\it Swift} J2058+0516 presented in this paper. 

In section 2 we discuss the data reduction and calibration of our optical imaging polarimetry of {\it Swift} J2058+0516, and in section 3 we compare our findings with those for {\it Swift} J1644+57 and recent numerical models.

\section{Observations and data analysis}

\subsection{VLT polarimetry} \label{sec:vltpol}

\begin{figure}
	\includegraphics[width=\columnwidth]{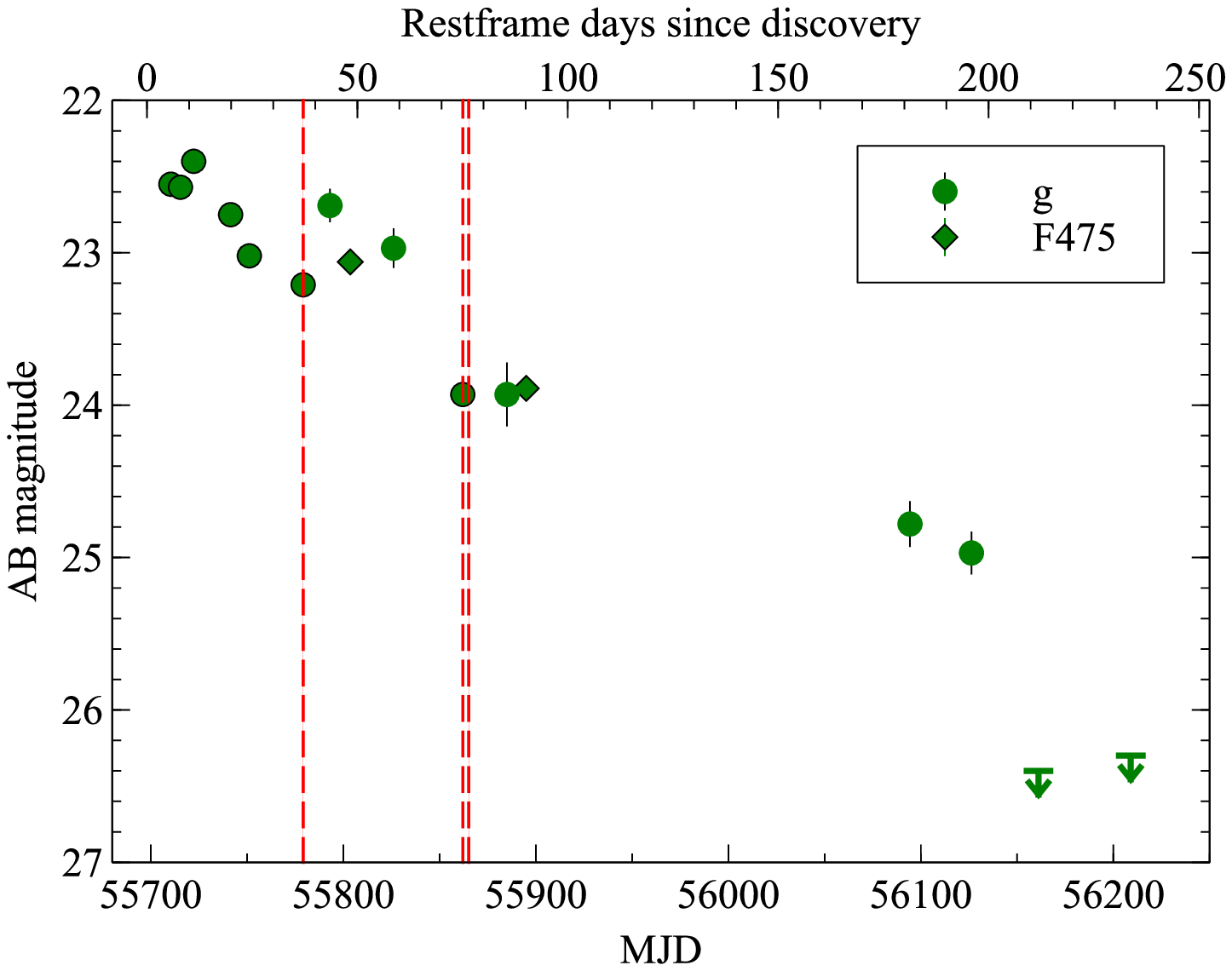}
    \caption{The optical lightcurve in $g$ band, from Cenko et al.~(2012)  and Pasham et al.~(2015), including data from Keck, the Gamma-Ray Burst Optical/Near-Infrared Detector (GROND), the William Herschel Telescope (WHT), Gemini-South, the Very Large Telescope (VLT) and the Hubble Space telescope (HST).
    The vertical lines indicate the epochs of polarimetry, magnitudes from the acquisition images of the polarimetry sequences are added to the plot; magnitudes are not corrected for Galactic extinction. A steady decay with some variability is seen, with a steep drop at late times (Pasham et al. 2015).}
    \label{fig:lc}
\end{figure}

We observed {\it Swift} J2058+0516 using the  FOcal Reducer/low dispersion Spectrograph 2 (FORS2) on the Very Large Telescope (VLT) of the European Southern Observatory (ESO), using its imaging polarimetry observing mode. Observations were obtained in Service mode. We acquired data at three epochs (Table \ref{tab:VLTobs}), with each epoch consisting of a series of three or four individual waveplate angle sequences. Each sequence consisted of four exposures that were taken with a Wollaston element and a half-wave plate in the beam. The half-wave plate was rotated at angles of $0^{\circ}$, $45^{\circ}$, $22.5^{\circ}$ and $67.5^{\circ}$ within one sequence. The Wollaston element splits the incoming light into two, the ordinary and extraordinary beams, which are orthogonally polarised. A strip mask is used to prevent overlap of the ordinary and extraordinary beam images. All polarimetric observations used the ESO $v\_{\rm HIGH}$ filter, and a $2\times2$ binning of the detector. We calculated the weighted average of all sequences per epoch to optimise our measurement signal to noise ratio. The midtimes of the three epochs, the number of sequences, exposure times and seeing conditions are listed in Table \ref{tab:VLTobs}. The data quality of the second and third epoch are not optimal: there was poor seeing in epoch 2, and a high sky background (because of proximity to the Moon) in epoch 3.

Acquisition images were obtained as part of the target acquisition procedure for the polarimetry observations, used to place the target into the strip mask. All acquisition images used the filter $v\_{\rm HIGH}$. After data reduction, the acquisition images belonging to a single polarimetry epoch were combined, the source was well detected in epoch 1 and epoch 2 acquisition data, but only marginally in epoch 3 (due to the poor observing conditions). We used aperture photometry to extract fluxes, and calibrated these directly onto 
Pan-STARRS\footnote{https://panstarrs.stsci.edu} $g$ band values for field stars (ignoring colour terms). The lightcurve, including the acquisition image photometry and the photometry from \cite{Cenko} and \cite{Pasham}, is shown in Figure \ref{fig:lc}, with vertical dashed lines indicating the epochs of polarimetry. The lightcurve appears to show some degree of variability on top of a steady decline, with a steep drop at late times
(\citealt{Pasham}).  Note that a wide range of instruments (and calibrations) were used, and one should be cautious in interpretation of apparent features in this  lightcurve.

The polarimetric data were reduced using routines in {\sc IRAF}\footnote{{\sc IRAF} (Image Reduction and Analysis Facility) is distributed by the National Optical Astronomy Observatories, which are operated by the Association of Universities for Research in Astronomy, Inc., under cooperative agreement with the National Science Foundation.}, following the procedures outlined in \cite{Wiersema121024}, using imaging flatfields and bias frames taken as close in time to the science data as possible. Fluxes for the TDE and field stars were extracted using aperture photometry in the same manner as in \cite{Wiersema121024}. 
We calculated the normalized Stokes parameters $q = Q/I$ and $u = U/I$ using a Python3\footnote{www.python.org} script following the method described in \cite{Patat2006}. These were then converted into the degree of linear polarisation and the polarisation angle, $P$ and $\theta$ respectively,  using the following relations
\begin{align}
\begin{split} \label{eq:linp}
P = \sqrt{q^{2} + u^{2}}
\end{split} \\
\begin{split} \label{eq:theta}
\theta = \frac{1}{2}{\rm arctan}\left(\frac{q}{u}\right) + \phi
\end{split} \\
\begin{split} \label{eq:phi}
\phi =
\begin{cases}
	0^{\circ},& \text{if } q > 0 \text{ and u} \geq 0\\
	180^{\circ},& \text{if } q > 0 \text{ and u} < 0\\
	90^{\circ},& \text{if } q < 0
\end{cases} 
\end{split}
\end{align}
where $\phi$ is an offset angle determined by the measured values of $q$ and $u$
(see \citealt{Wiersema091018}). Note that we assume there is no polarimetric evolution within one epoch (i.e. we combine the sequences for each epoch together).

The measured linear polarisation has to be corrected for polarisation bias. The bias arises as linear polarisation is derived from the addition of $q$ and $u$ in quadrature (see Equation \ref{eq:linp} or \citealt{Serkowski1958}). There are a number of estimators that can correct for this bias, dependent on the signal to noise ratio (SNR; in flux as well as polarisation) of the source (e.g.  \citealt{Simmons1985}). We use the modified asymptotic (MAS) estimator described in \citet{Plaszczynski2014} to correct for the polarisation bias. The generalised estimator is defined as follows
\begin{equation} \label{eq:pmas}
P_{\rm MAS} = P - \sigma^{2} \left[\frac{1-e^{\frac{-P^2}{\sigma^{2}}}}{2P}\right] 
\end{equation}
where $P_{\rm MAS}$ is the estimation of the true polarisation $P_{\rm 0}$ and $\sigma$ is the standard error on $P$. The variance on $P$ can be found using the following relation
\begin{equation} \label{eq:sterr}
\sigma^{2} = \frac{u^{2}\sigma^{2}_{u} + q^{2}\sigma^{2}_{q}}{q^{2}+u^{2}}
\end{equation}
where $\sigma_{q}$ and $\sigma_{u}$ are the standard errors on $q$ and $u$, respectively. Equation \ref{eq:sterr} assumes that $q$ and $u$ are independent.

For epochs where we have a positive detection of polarisation (i.e. $P/\sigma\gtrsim3$) the distribution of $P$ is approximately Gaussian. Therefore, for our second epoch we can simply quote the $1\sigma$ errors. As the signal to noise ratio  decreases, the distribution of $P$ no longer follows a Gaussian distribution but instead follows a Rice distribution (\citealt{Rice1944}; see also \citealt{Patat2006} for a numerical demonstration). This transition results in more complex confidence intervals for lower signal to noise ratios. We can calculate an upper limit on $P$ using the following analytical relation from \cite{Plaszczynski2014}
\begin{align}
P^{\alpha}_{\rm Upper} = P_{\rm MAS} + P_{\alpha}(1 - \beta e^{-\gamma P_{\rm MAS}})
\end{align}
where $\alpha = 0.95$, $P_{\alpha} = 1.95\sigma$, $\beta = 0.22$ and $\gamma = 2.54$ for a $2\sigma$ upper limit - which we quote for our first and third epochs. The measurements for all three epochs are shown in Table \ref{tab:VLTobs}.  

\begin{table*}
	\centering
	\caption{Log of VLT polarimetry observations of {\it Swift} J2058+0516, uncertainties are 1$\sigma$, upper limits are given at 2$\sigma$ level (see Section \ref{sec:vltpol}). $^{a}$: A sequence is defined as a set of exposures at four wave plate angles ($0^{\circ}$, $45^{\circ}$, $22.5^{\circ}$ and $67.5^{\circ}$). $^{b}$: Exposure time per wave plate angle per sequence.
    $^{c}$: From acquisition images, calibrated onto Pan-STARRS $g$ band values of field stars (\citealt{Flewelling}). The data quality of the acquisition image of the third epoch was too poor to measure a reliable magnitude. The discovery of the source by the Burst Alert Telescope (BAT) on {\it Swift} was on or around 17 May 2011 (MJD 55698), see Cenko et al.~(2012) for details.
    }
	\label{tab:VLTobs}
	\begin{tabular}{lllllllll} 
		\hline
		MJD Date & Sequences$^{a}$  & Exp. time$^{b}$ & TDE magnitude$^{c}$& Seeing & $q$(x$100\%$) & $u$(x$100\%$) & $P$(x$100\%$) & $\theta$ \\
		       &                   & (s)       &     (AB)            & (arcsec) & & & & ($^\circ$) \\
        \hline
         55779.14848        & 3    & 300      &    $23.21\pm0.05$     & 0.8 & 2.20($\pm1.86$) & -1.36($\pm1.04$) & $<5.25$ & - \\ 
         55862.07378        & 4    & 335      &    $23.93 \pm 0.07$   & 1.5 & 7.26($\pm2.16$) & 4.40($\pm3.31$) & 8.12($\pm2.52$) & 15.6($\pm8.51$) \\  
         55865.07486        & 4    & 335      &                      & 0.9 & 1.26($\pm1.59$) & -5.68($\pm3.29$) & $<12.88$ & - \\
     \hline
	\end{tabular}
\end{table*}

\subsection{Off axis instrumental polarisation/ line-of-sight Galactic dust contribution}
The intrinsic polarisation of {\it Swift} J2058+0516 is not the only contributor to our measured polarisation. Dust particles residing on our line-of-sight can induce a significant polarisation signature through scattering. We attempted to quantify the Galactic dust contribution using field stars within the VLT field-of-view, assuming that the average intrinsic polarisation of field stars is zero.

\begin{figure}
\includegraphics[width=\linewidth]{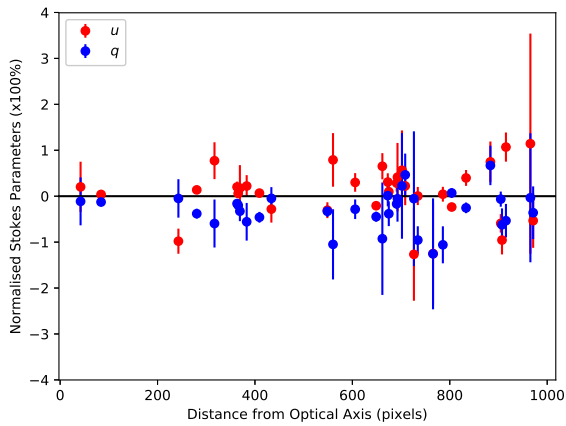}
\caption{Plot representing the apparent normalised Stokes $q$ and $u$ polarisations of the unpolarised field stars away from the optical axis. \label{fig:qudist}}
\end{figure}

\begin{figure}
\includegraphics[width=\linewidth]{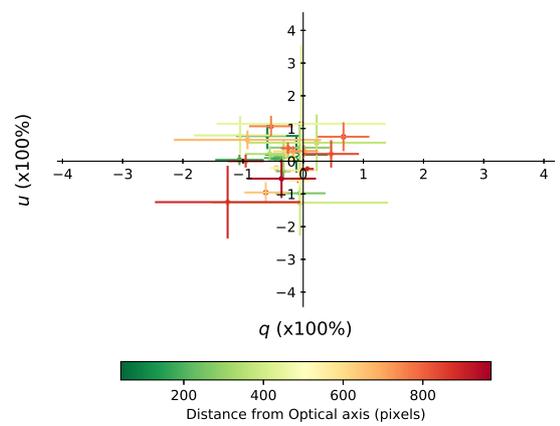}
\caption{Plot representing the apparent polarisation of unpolarised field stars in normalised Stokes $q$ and $u$ parameter space. This is caused due to the presence of instrumental polarisation away from the optical axis. The colours represent the distance away from the optical axis.\label{fig:qufieldstars}}
\end{figure}

Our analysis included a total of 34 field stars at varying radial distances from the optical axis (where the TDE is placed). We calculated the $q$ and $u$ values utilising the weighted average from all 11 observations over the three epochs epoch to optimise our signal to noise ratio - as the stars are intrinsically unpolarised the polarisation signature in each epoch should be identical. The field stars are homogeneously distributed spatially around the optical axis and so we are confident that the summation of the contribution of instrumental polarisation from all of these sources is negligible (the instrumental polarisation likely has a weak radial pattern, increasing with radial distance from the optical axis, see e.g. \citealt{Patat2006} and Gonz\'{a}lez et al.~in prep.). The field stars chosen covered a range of magnitudes from $\sim 18-22.5$\,mag in the AB system. We opted for brighter sources to reduce the uncertainties on individual source measurements. Any measured offset in the centre of the Stokes $q$ and $u$ distribution should therefore come from the Galactic dust contribution. We calculated the polarisation signature using the same method described in section \ref{sec:vltpol}, 
Figures \ref{fig:qudist} and \ref{fig:qufieldstars} show the field star $q,u$ distribution. 
The resulting estimate for the Galactic line of sight contribution of the polarisation is  displayed in table \ref{tab:dustparams}: we find a low value of $P<0.48\%$, in line with expectations from the low line of sight Galactic dust extinction, $E(B - V) \approx 0.095$ mag \citep{SchlaflyFinkbeiner}.

\begin{figure}
\includegraphics[width=\linewidth]{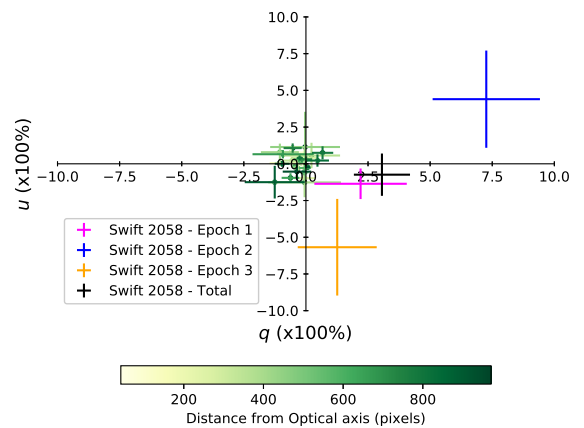}\label{fig:qu2058}
\caption{Plot showing the normalised Stokes $q$ and $u$ measurements for the three observational epochs of {\it Swift} J2058+0516 and the combined $q$ and $u$ measurements. The $q$ and $u$ values of the field stars are over plotted for comparison.}
\end{figure}

\begin{table}
	\centering
	\caption{Estimation of the $q$, $u$ and $P$ Stokes parameters induced by scattering by Galactic dust in the line of sight.}
	\label{tab:dustparams}
	\begin{tabular}{lll}
    \hline
    $q$ (x$100\%$) & $u$ (x$100\%$) & $P$ (x$100\%$) \\ \hline
    -0.29($\pm0.10$) & 0.11($\pm0.15$) & <0.48 \\ \hline
    \end{tabular}   
\end{table}

\section{Discussion}
The measurements shown in Table \ref{tab:VLTobs} show a non-detection at epoch 1, a detection at epoch 2 ($P = 8.1 \pm 2.5 \%$), and a non-detection at epoch 3. Epochs 2 and 3 are taken very close to each other, but conditions at epoch 3 were considerably poorer, and the 2$\sigma$ limit for epoch 3 is consistent with the detected level in epoch 2. At epoch 1 we see a non-detection, with a 2$\sigma$ limit of 5.3\%. This may imply some variation in $P$ between epochs 1 and 2, which is more clearly seen in the $q,u$ plane (Figure 4), where epoch 2 and 3 appear more distinct. However, given the relatively large uncertainties, a constant level of $P$ of $\sim8\%$ (the level detected in epoch 2) can not be confidently ruled out. We further note that we assumed there is no polarisation variations at timescales of a single observation epoch. If the polarisation angle changes rotates fast compared to the duration of an observing epoch, the polarisation would be smeared out and variation difficult to measure.

\subsection{Comparison with {\it Swift} J1644+57}
The only other jetted TDE with linear polarimetry measurements is {\it Swift} J1644+57. This source showed a non-zero linear polarisation in the $K_s$ band, of $P_{\rm K_s} \sim7.4\pm3.5 \%$, measured using the William Herschel Telescope (\citealt{Wiersema1644}). At first glance, the measured linear polarisation of {\it Swift} J1644+57 and {\it Swift} J2058+0516 are remarkably similar, however there are some caveats to this comparison. First of all, the very high levels of dust extinction in the case of {\it Swift} J1644+57 likely imply a non-negligible contribution by dust scattering to the  
observed linear polarisation, though it is very unlikely that all the polarisation is caused by dust scattering in the host (\citealt{Wiersema1644}). 
Secondly, the observations of {\it Swift} J1644+57 were done in the near-infrared $K_s$ band and the source redshift is $z = 0.3543$ (\citealt{Levan}), the observations of {\it Swift} J2058+0516 were done in the $v_{\rm HIGH}$ filter and the source redshift is $z = 1.1853$ (\citealt{Cenko}). This means that for {\it Swift} J1644+57 the polarimetry was done in the rest-frame near-infrared, whereas for {\it Swift} J2058+0516 these were done in the rest-frame near-ultraviolet. These wavelength regimes may have somewhat different origins (\citealt{Curd}; see section \ref{sec:models}).
Thirdly, the observations of {\it Swift} J1644+57 are obtained when the source showed a plateau-like lightcurve evolution (Figure 3 in \citealt{Wiersema1644}), whereas {\it Swift} J2058+0516 showed a somewhat more steady decline (\citealt{Pasham}).

\subsection{Radio polarimetry}
The relativistic jet physics of TDEs is arguably easiest to study at radio wavelengths, where the forward shock from the jet is bright.
Indeed a rich phenomenology is seen that can be followed from just hours after the jet launch to years after. Abundant radio observations over a wide radio spectrum were obtained for {\it Swift} J1644$+$57 (\citealt{Zauderer}; \citealt{Berger};  \citealt{Wiersema1644}; \citealt{Zauderer2}; \citealt{Cendes}; \citealt{Yang}; \citealt{Eftekhari}).
For {\it Swift} J2058$+$0516, the data-set at radio wavelengths is considerably smaller (\citealt{Cenko}; \citealt{Pasham}) in terms of lightcurve and spectral coverage.
In \cite{Wiersema1644} we derived deep limits on the polarisation of the radio emisison of {\it Swift} J1644+57 (most sensitive 3$\sigma$ limits as deep as 2.1\%), using deep observations with the  Westerbork Synthesis Radio Telescope, at a range of wavelengths and timescales. To date, this remains the only published radio polarimetry data of a TDE. A small number of flux values at radio wavelengths for {\it Swift} J2058+0516 have been reported in the literature (\citealt{Cenko}; \citealt{Pasham}),
using the Very Large Array (VLA) and the Very Long Baseline Array (VLBA), but we are unable to derive sensitive polarisation measurements or limits for these observations, as insufficient polarisation calibrator observations were obtained.

\subsection{Comparison with models}\label{sec:models}
It appears likely that the diversity in TDE emission is related to the dynamics and geometry of the accretion flow (with as key parameters the strength of the magnetic field around the black hole, the black hole mass and the black hole spin parameter), the accretion rate, and the viewing angle. Optical emission may originate from a variety of mechanisms, e.g. from  the accretion disk, the jet, reprocessed X-ray emission, or from shocks generated by stellar debris self-crossing \citep{Piran}.

\cite{Dai} and \cite{Curd} present a set of 3 dimensional general relativistic magneto-hydrodynamic simulations to explore the impact of the various parameters on the resulting TDE spectra and lightcurves. Both groups find a relativistic jet is generated in some of their models. 
In particular, \cite{Curd} find that one of their models has a clear collimated relativistic outflow, with properties broadly consistent with those seen in jetted TDEs. 
The spectra computed by \cite{Curd} for the cases that do not produce a jet, match fairly well with observations of non-jetted TDEs, in particular reproducing the thermal components and properties of the X-ray emission. 
There are some discrepancies with observations and with the work from \cite{Dai} regarding the X-ray and optical emission for sources at different inclination angles.
\cite{Curd} compare their simulation with data of {\it Swift} J1644+57, finding reasonable agreement with X-ray observations and jet structure,
but the large, and uncertain, extinction in the line of sight of {\it Swift} J1644+57 precludes a comparison in the UV and optical range.
Their model predictions indicate the presence of strong, beamed, X-ray and $\gamma$-ray emission from the jet. In the ultraviolet a thermal peak is predicted arising from the torus, with in the far infrared a thermal synchrotron peak. In the optical domain the received light comes in part from the torus and in part from the outflow in the models of \cite{Curd}, in proportions depending on the viewing angle.

An alternative model of TDE accretion and emission is presented by \cite{Coughlin}. They argue that angular momentum of the infalling matter is too small to produce an accretion disk. Instead, the accretion energy is trapped and inflates the infalling gas to form a quasi-spherical ``ZEro-BeRnoulli Accretion" flow, or ZEBRA. ZEBRAs radiate as blackbodies, but can only do so up to the Eddington limit. Any additional accretion energy, injected into the inner regions of the ZEBRA by the black hole, is also unable to be efficiently advected to the surface inhibiting wind formation. Instead, this excess energy must escape as polar jets where the surface of the ZEBRA envelope can be exited. This model therefore accounts for both jetted and non-jetted TDEs as dependent on the observation angle, similarly to the unified model proposed by \citet{Dai}. \textit{Swift} J1644$+$57, \textit{Swift} J1112.2$-$8238 and \textit{Swift} J2058$+$0516 have all previously been examined in relation to the ZEBRA model (\citealt{Coughlin}; \citealt{Wu}). In particular, the ZEBRA model demonstrates a good agreement with the X-ray data for all three events, specifically the time scale at which the X-ray flux drops as the accretion rate falls to sub-Eddington levels and the jet turns off, while the power law spectrum and luminosity of \textit{Swift} J1644$+$57's jet is consistent with the predictions from the ZEBRA model. During the jetted phase of \textit{Swift} J2058$+$0516, the temperature of the thermal-like  spectrum observed in the optical and UV is also consistent with the ZEBRA model, particularly with a black hole mass of $\sim5\times 10^6$ solar masses, within the constraints found by \cite{Cenko}. Comparing the ZEBRA model to non-jetted TDEs, \cite{Wu} find that the temperatures of $\sim$ a few $\times 10^4$ K generally predicted are also more consistent with observations. In particular, there is a strong agreement with the SEDs of iPTF16axa (\citealt{Hung2}) and PS1-10jh (\citealt{Gezari}).

We may expect significant linear polarisation for (optical)  emission from the jet and for (inverse) Compton scattered emission, but much less so for the thermal torus emission. The forward shock emission (the interaction of the jet with external medium, not analysed in e.g. \citealt{Curd}) will be linearly polarised as well, with the amount depending on the jet structure (which is likely not homogeneous for at least {\it Swift} J1644+57, see e.g. \citealt{Mimica}), the viewing angle and the jet opening angle, and the detailed properties of the magnetic field in the shock, i.e. is all the field random and confined to the shock, or is there an ordered component perpendicular to the shock - this is a well studied problem in gamma-ray burst afterglows (e.g. \citealt{Gill}). At longer wavelengths, plasma propagation effects can give rise to significant depolarisation of the forward shock radio emission (e.g. \citealt{Toma}). In addition to these internal effects, we are also likely to see the effect of dust scattering on the received polarisation (see discussion in \citealt{Wiersema1644}), which can polarise intrinsically unpolarised emission, convert linear to circular polarisation (see discussion in \citealt{Wiersema121024}), and alter the linear polarisation properties of intrisically polarised light. In the case of {\it Swift} J1644+57 there is clear evidence of a large dust column, based on the very red colours of the transient light, whereas for {\it Swift} J2058+0516 the inferred amount of dust in the line of sight is far lower: \cite{Pasham} place a limit of
$A_V\lesssim0.2$ mag (assuming a Milky Way like extinction law). This limit indicates that dust scattering along the line of sight is unlikely to contribute more than $\sim1\%$ polarisation at the observed wavelength (a more exact limit requires a better understanding of the dust grain size distribution).
Our detection of linear polarisation in {\it Swift} J2058+0516 at rest-frame blue optical/near-UV range indicates that some non-thermal emission is present in the received flux. The polarisation $P$ is moderate, and may be in agreement with the predicted flux 
origins by \cite{Curd}, though a more quantitative comparison would require polarisation measurements of a larger number of sources, at a larger number of wavelengths, and with smaller uncertainties.

\subsection{Future TDE polarimetric follow-up}
The sample of polarimetrically studied TDEs is very small. The predicted future TDE yield of large (optical) surveys like ZTF and LSST is large
(\citealt{Hung}; \citealt{LSST}), and will allow a more systematic approach to TDE polarimetry in both optical and longer wavelengths. In \cite{Higgins} we trialled 
a linear polarimetry programme (the Snapshot survey for Polarised Light in Optical Transients, SPLOT) that obtained snapshot optical linear polarimetry of a large sample of randomly selected optical transients, which included a thermal TDE. A similar approach, using high-volume optical transient streams like the ZTF transient stream, and pre-selecting on nuclear transients, may greatly increase the yield of TDE polarimetry. It would simultaneously probe the effect of dust scattering in a statistical manner. 
A targeted optical polarimetric survey of TDE candidates with a high likelihood of having a relativistic jet, i.e. ones triggered by space based X-ray or $\gamma$-ray detectors, or found in wide-field radio surveys (e.g. by MeerKAT; \citealt{Booth}), can further elucidate the origin of the different spectral components of TDEs. Optical and near-infrared spectro-polarimetry would be a key extension to the imaging polarimetry presented in this paper and in \cite{Wiersema1644}. At the same time, a sample of thermal TDEs could remain valuable as a comparison, and has yet to be obtained. Since thermal TDEs are more common, and hence often at lower redshift, such observations can 
be obtained more readily. 
While limited to the brightest subsample ({\it Swift} J2058+0516 was too faint for optical spectro-polarimetry throughout its evolution), the effects of dust scattering and the origin of the non-thermal emission have different wavelength dependency, so spectro-polarimetry would easily allow breaking of degeneracies, in a manner similar to spectro-polarimetric studies of (core-collapse) supernovae. 

In the X-ray regime, TDEs are an important component of the science cases for future X-ray polarimetry-capable missions such as IXPE and eXTP (e.g. \citealt{intzand_eXTP}). The jetted TDEs are very bright at X-ray wavelengths, giving X-ray polarimetry with small statistical errors even at relatively short exposure times (\citealt{intzand_eXTP}), allowing tests for jet precession, the formation of globally ordered fields, and variations of the magnetic field over short timescales (\citealt{intzand_eXTP}). 
At radio  wavelengths, the advent of large, wide-field surveys (e.g. with the WSRT Apertif, \citealt{Oosterloo}) will provide a way to select jetted TDEs via their non-thermal radio emission, and collect polarisation information for a large sample (Apertif will survey the northern sky in polarised continuum). Deep circular radio polarimetry will similarly help test models for Faraday conversion in the jet, or in the intervening medium. Ultimately, the success of polarimetry for jetted-TDEs will depend both the availabilty of suitable resources, and
on the detection of candidates, which to date, have been rare. 

\section{Conclusions}
In this paper, we present optical ($V$ band) imaging polarimetry of {\it Swift} J2058+0516 using the VLT. This is only the second jet-driving, relativistic, tidal disruption event studied using polarimetry, after {\it Swift} J1644+57. We obtained three epochs of data, in which we find evidence of linear polarsation in the second epoch, at a level $P = 8.1 \pm 2.5 \%$, the other two epochs provide upper limits. There is weak evidence for polarisation variation between the epochs. We compare the polarisation information to current basic models, and find that these can accomodate small levels of linear polarisation as measured. Our measurements of two relativistic TDEs show the value of linear polarimetry as a tool to better understand the contributions of disk, jet and winds to the received spectrum. Polarimetry over a wider wavelength range will help to break existing degeneracies in comparison of models with lightcurve and broadband spectral energy distributions.

\section*{Acknowledgements}
We thank the anonymous referee for their constructive comments that improved this paper.
Based on observations collected at the European Organisation for Astronomical Research in the Southern Hemisphere under ESO programme 288.B-5011(A), PI A. Levan. KW, AJL \& NRT acknowledge that this project has received funding from the European Research Council (ERC) under the European Union$’$s Horizon 2020 research and innovation programme (grant agreement no 725246). 
NRT, RLCS, ABH and RAJE thank the STFC for support. RLCS acknowledges support from Royal Society Research Grant RG170230. KW thanks S. Covino for insightful discussion.











\bsp	
\label{lastpage}
\end{document}